\documentclass[prd,showpacs]{revtex4}
\usepackage[dvipdfm]{hyperref}
\begin{document}
\title{A Criterion of Naturalness in Renormalization}

\author{Su Yan}
\email{yans@northwestern.edu}
\affiliation{Department of Physics and Astronomy,
Northwestern University, Evanston, IL 60201}

\begin{abstract}
The sensitivity criterion is widely used in measuring the level of 
fine-tuning, although many examples show it doesn't work under certain circumstances. 
We discuss the mathematics behind the fine-tuning problems, explain the mathematical meanings of the 
sensitivity criterion, point out three implicit assumptions behind this criterion.
Because of these assumptions, the sensitivity criterion can't reflect the fine-tuning level correctly.
By analyzing two well known examples that the sensitivity criterion failed, 
we point out the dimensional effect is the main reason why we have these problems.
To solve these problems, we  proposed a new criterion to replace the sensitivity criterion.
\end{abstract}
 

\maketitle
\section{Introduction: The Naturalness Problem}
The principle of naturalness introduced by Wilson and 't Hooft\cite{Wilson} requires 
that in order to obtain a weak scale observable parameter of the order \(M_W\),  we do not 
need to extremely fine-tune the fundamental Lagrangian parameters at the grand 
unification scale. Although we have different fine-tuning problems in different models, 
generally, these problems can be categorized into two types:
The first type fine-tuning problems, which existed in renormalization.  
The second type fine-tuning problems, which existed in the
mixing mechanisms or when parameters are linked together by matrix transformation. 

The first type fine-tuning problems, 
for example, the renormalization of \(\phi^4\) model:

\begin{equation}
\mathcal{L}=\frac{1}{2}[(\partial_{\mu}\phi)^2-m^2_0\phi^2]-\frac{g}{4!}\phi^4
\label{eqx1}
\end{equation}
the renormalized scalar mass \(m^2\) is:
\begin{equation}
m^2=m^2_0-g^2\Lambda^2
\label{eqx2}
\end{equation}
where \(m_0\) is the bare mass, 
\(\Lambda\) is the cut-off energy scale. Because both \(m_0\) and \(\Lambda\)
are around \(10^{18}\) GeV, thus to obtain a small weak scale 
renormalized mass \(m\), we need a fine-tuning mechanism to match the values of \(m_0\) and \(\Lambda\).

For the first type fine-tuning problems, generally,  the parameter 
range at the weak scale(which is 0 to \(M_W\) roughly)
is much smaller than the corresponding  parameter range at the grand unification scale
(which is 0 to \(M_P\) roughly). 
The parameter range shrinks rapidly as the energy scale decreases. Here
the fine-tuning problem is due to the large parameter range shrinkage from the grand unification
scale to the weak scale (This will be more clearly if it is expressed in a renormalization group
equation). It happens when we compare parameters with different energy scales,
and once we given the value of the weak scale parameters, the large parameter range at the grand 
unification scale will be controlled by the renormalization group equation itself.

The second type fine-tuning problems generally is not related with renormalization,
they are related with the parameter transformation at the same energy scale, 
for example, by mass mixing.
Take  the MSSM tree level Z-boson mass \(m_Z\) as an example:

\begin{equation}
\frac{1}{2}M^2_Z=\frac{M^2_{H_D}-M^2_{H_U}\tan^2\beta}{\tan^2\beta-1}-\mu^2
\label{eqx3}
\end{equation}

Because both \(M_{H_D}\), \(M_{H_U}\) and \(\mu\) could be as large as \(10^{5}\) GeV~\cite{Arkani-Hamed:2004fb}
,
while \(M_Z\) is only around 100 GeV, again we need a fine-tuning mechanism to  
match the values of \(M_{H_U}\) and \(\mu\). Besides this example, We will find this types of fine-tuning problems  
in the various  mass mixing mechanisms in the MSSM model. 

The example  
of the MSSM model Z-boson mass fine-tuning problem shows, for this type fine-tuning problems,
the relation between the input parameters (\(M^2_{H_D}\), \(M^2_{H_U}\) and \(\mu^2\))
and the output parameter (\(M^2_Z\)) is linear (or not far away from linear for other
similar examples in mass mixing mechanisms), which means
not like in the first type where the parameter range shrinks \(10^{18}\) times,  
the possible range of the output parameter \(M^2_Z\) is almost the same as the possible range of 
the input parameters \(M^2_{H_D}\), \(M^2_{H_U}\) and \(\mu^2\). Thus
the fine-tuning is mostly due to the possibility is extremely low to pick up a specific
parameter value in an extremely large parameter range. For this type, 
the large parameter range is not decided by the mixing mechanism or the transformation itself, 
it is given by other mechanisms. 

These two types of fine-tuning problems have different properties, we may need different
methods to describe them.

\section{The Sensitivity Criterion and its problems}
The sensitivity criterion proposed by R. Barbirei and G.F.Giudice et al.\cite{BG}
is the first widely adopted criterion in this field, it uses the sensitivity parameter to quantitatively 
measure the naturalness level.
If \(x\) is a fundamental Lagrangian parameter at the grand unification scale,
and \(y\) is a computed observable  parameter such as weak scale masses, Yukawa couplings.
If we varies the Lagrangian parameter \(x\) at the grand unification scale,  based on 
the corresponding variation of the  weak scale observable parameter \(y\), 
the sensitivity parameter \(c\) is defined as:

\begin{equation}
c(x_0)=\bigg| \frac{x}{y}\frac{\partial y}{\partial x}\bigg|_{x=x_0}
\label{eqx4}
\end{equation} 
here need to emphasis that  the Lagrangian parameter \(x\) and the 
observable parameter \(y\) may have different units, while the sensitivity parameter \(c\)
will be used to compare the fine-tuning properties of different models.
We would like to see it in a dimensionless formation. So in the definition of the sensitivity parameter 
\(c\), relative variations \(\delta y/y\) and \(\delta x/x\) are used as the basis of the comparison. 

Barbieri and Giudice chose \(c=10\) as the maximum allowed sensitivity for any
models to be categorized as ``natural'', 
if \(c\gg 10\) then it is ``unnatural'' (or fine-tuned). The sensitivity criterion 
has been widely adopted since then.  It 
has been used in many fields, for example,  setting a 
naturalness contour for SUSY particle search\cite{Azuelos:2002qw},
or the fine-tuning problem of the neutrino
seesaw mechanism\cite{Casas:2004gh}. But people soon found it is not a reliable criterion. Many examples
show that sometimes the  sensitivity criterion fails. The most 
famous examples among them are given by G. Anderson et al\cite{GWA} and P. Ciafaloni et al\cite{CS}.

The example given by  G. Anderson et al\cite{GWA} is regarding the high sensitivity
of \(\Lambda_{QCD}\) scale to the variation of the strong coupling constant \(g\).
Because the relation between \(\Lambda_{QCD}\) and the strong coupling constant \(g\) is:

\begin{equation}
\Lambda_{QCD}=M_P\exp{(-\frac{(4\pi)^2}{bg^2(M_P)})}
\label{eqx5}
\end{equation}
the corresponding sensitivity parameter \(c\) is:
\begin{equation}
c(g)=\frac{4\pi}{b}\frac{1}{\alpha_s(M_P)}\gtrsim 100
\label{eqx6}
\end{equation}
according to the sensitivity criterion, it is fine-tuned. But we know actually \(\Lambda_{QCD}\)
is protected by the gauge symmetry, it is not fine tuned.

The example given by P. Ciafaloni et al\cite{CS} is about the high sensitivity of 
the Z-boson mass. When the Z-boson mass 
\(M_Z\) is dynamically determined through gaugino condensation in a ``hidden'' sector, 
the mass \(M_Z\) can be expressed as:
\begin{equation}
M_Z\approx M_P e^{-l/g^2_H}
\label{eqx7}
\end{equation}
Where \(g_H\) is the hidden sector gauge coupling constant renormalized at \(M_p\),
and \(l\) is a constant.  Like the first example, the calculated sensitivity \(c\) for 
the second example is also much larger than \(10\), the maximum allowed sensitivity, 
although we know Z-boson mass is also not fine-tuned.

Compare Eq.~(\ref{eqx5}) and Eq.~(\ref{eqx7}), both of them have a similar
mathematical formation.
Actually, if we calculate the sensitivity \(c\) of any weak scale mass 
by varying a related coupling constant at the grand unification scale, the calculated sensitivity
is always very 
high no matter whether the mass is really
fine-tuned or not. These examples show the sensitivity criterion is not reliable.
Although this problem has been pointed out long time ago, 
yet we still can find people use the grand unification scale coupling constant 
to measure the level of the fine-tuning\cite{Casas:2004gh}.

Besides these well known problems, the sensitivity criterion also has other problems.
If we apply the sensitivity criterion 
(Eq.~(\ref{eqx4})) to the scalar mass of  \(\phi^4\) model (Eq.~(\ref{eqx1})),
clearly from Eq.~(\ref{eqx2}) we know the scalar mass is highly fine-tuned. 
But on the other hand, because both \(m^2_0\) and \(g^2\Lambda^2\) in Eq.~(\ref{eqx1}) are not 
independent, so if we integrate the mass renormalization group equation, we have:

\begin{equation}
m^2=m^2_0 e^{\int_0^t(g^2/16\pi^2-1)dt}
\label{eqx13}
\end{equation}

Thus the sensitivity \(|\partial\ln m^2/\partial\ln m^2_0|\) equals to one, and it is not fine-tuned.

Obviously, the origin of this problem is the ratio \(\delta m^2/m^2\) is not depended on 
the energy scale. Thus
even \(\delta m^2\) is only around hundreds \(\mbox{GeV}^2\) while \(m^2\) is 
around  \(10^{36} \mbox{GeV}^2\)(which means highly fine-tuned), the sensitivity is still one.
This reminds us the sensitivity \(c\) can not reflect the level of the fine-tuning 
when it is due to the shrinkage of the parameter range. 

Generally,  for renormalization, if we integrate a renormalization group equation,
roughly speaking, an observable  parameter \(y\) can be expressed as a function of the energy scale:
\begin{equation}
y=y_0 e^{\int(n-\gamma(y_0))dt}
\end{equation}
where \(n\) is the naive dimension and \(\gamma\) is the anomalous dimension.
If the naive dimension \(n\) is not equal to zero,  for example, \(\phi^4\) model scalar 
mass, then there will be a fine-tuning problem. If we calculate the sensitivity \(c\) of 
the weak scale parameter \(y\) when we varies \(y_0\) at the grand unification
scale,  the corresponding sensitivity \(c\) is:
\begin{equation}
c(y_0)=1-\frac{\partial}{\partial\ln y_0}\int \gamma(y_0)dt
\end{equation}
Which means the sensitivity \(c\) can only reflect the contribution from 
the anomalous dimension, 
and will ignore the major contribution from the big naive dimension.
So in some sense it is more accurate to call the sensitivity \(c\) as the anomalous sensitivity
when dealing with the first type fine-tuning problems. 

Based on the above analysis, we learn that the sensitivity criterion has many problems.
First it can not be applied to the first type fine-tuning problems, and the second, 
it is incorrect when dealing with parameters with different dimensions.
To solve these problems, many authors proposed alternative prescriptions.
G. Anderson et al~\cite{GWA,AC2,AC3}. first introduced the idea of probability
distribution, they argued that, the relative variation could have a probability 
distribution,  we need to use \(\bar{c}\), the probability average of the 
sensitivity \(c\), and the sensitivity criterion should be replaced by:
\begin{equation}
\gamma = c/\bar{c}
\label{eqx8}
\end{equation}
Based on this criterion, only those with \(\gamma \gg 1\) should be considered as fine-tuned.

Later another modified definition was proposed to replace the original sensitivity criterion
~\cite{CS,BR,BS,GRS,RS}:

\begin{equation}
c(x_0)=\bigg| \frac{\Delta x}{y}\frac{\partial y}{\partial x}\bigg|_{x=x_0}
\label{eqx9}
\end{equation} 
where \(\Delta y\) is the experimentally allowed range of the parameter \(x\).

Although many authors attempted to give a correct numerical description of the naturalness
level, none of them can claim quantitative rigor. 
The calculated naturalness level usually depends on what criterion we used,
and these results may reflect the naturalness properties correctly or incorrectly.
Because the sensitivity criterion plays a very important role, it is worth to investigate 
the relationship between the naturalness and the sensitivity, thus we can find a correct 
and reliable criterion.

\section{Mathematical basis of the naturalness criterion}
 
The naturalness problem is similar to the  
initial condition sensitivity problem in mathematics, which
means small differences in initial conditions will
result in big differences as the system  evolves. 
The final state of a system  will extremely depend on  small variations of the 
initial condition. In mathematics, sensitivity to initial conditions can be measured 
quantitatively in many ways, for example, by the Lyapunov exponent. These methods are 
all based on the probability theory, which reflects how large the probability is  
to find an output parameter to be in a
certain parameter range.

According to the probability theory, 
if parameters \(x\) and \(y\) are linked by
 \(y=f(x)\), and if  \(p[x]\) is the original 
probability distribution function of the parameter \(x\),
and if the function \(f(x)\) is monotonic, 
then the probability distribution of the parameter \(y\) will be~\cite{PR}:
\begin{equation}
p(y)=p[f^{-1}(y)]\bigg|\frac{\partial}{\partial y}f^{-1}(y)\bigg|_{x=x_0}
\label{eqx10}
\end{equation}
Note the probability distribution \(p(y)\) is controlled by the 
original distribution of the parameter \(x\)
(which is \(p[f^{-1}(y)]\)) and the transformation between the parameters \(x\) and \(y\)
(which is \(|\partial x/\partial y|\)). 

By means of the probability theory, we can interpret  the sensitivity criterion more clearly.
If we calculate the sensitivity of a parameter \(y\) by varying \(x\) around \(x=x_0\),
if we assume that the input parameter \(x\) is uniformly distributed from \(0\) to \(x_0\)
(probability density  \(p[f^{-1}(y)]=1/x_0\)),
then the total probability of the output parameter \(y\) to be found in the region \(0\le y\le y_0\) is:
\begin{equation}
P= \int_0^{y_0}\frac{y_0}{x_0} \bigg| \frac{\partial x}{\partial y}\bigg| dy\approx 
\frac{y_0}{x_0} \bigg| \frac{\partial x}{\partial y}\bigg|_{x=x_0}
\label{eqx11}
\end{equation}
Where \(P\) represents the total probability of the parameter \(y\) to be found within \(0\le y \le y_0\). 
In the second part of Eq.~(\ref{eqx11}) it is assumed that
\(\partial x/\partial y\) is a constant in the region \(0\le y\le y_0\).
Compare Eq.~(\ref{eqx11}) with Eq.~(\ref{eqx4}), we found that, if the total probability of \(y\)
to be found in the region  \(0\le y \le y_0\) is \(P\),  then 
the sensitivity parameter \(c\)  is approximately the inverse of this probability:
\begin{equation}
c=\frac{1}{P}
\label{eqx12}
\end{equation}

If we choose \(c_{\mbox{max}}=10\) as the maximum allowed sensitivity, by means of the 
probability theory, it means the minimum allowed total probability of the observable parameter
\(y\) to be found within the range \([0,y_0]\) is approximately \(10\%\).

Based on the above analysis, it is not difficult to see there are three assumptions for the sensitivity criterion:
\begin{enumerate}
\item The sensitivity criterion implies the Lagrangian parameter \(x\) is evenly distributed in
   \(0 \le x \le x_0\) only. The probability distribution in the region \(x > 0\) is zero.
\item What the sensitivity criterion calculated is approximately the \textit{total probability} of the observable
parameter \(y\) to be found  in the region \(0\le y \le y_0\).
\item The sensitivity criterion also implies that the relation between the grand unification 
scale parameter and the weak scale parameter is monotonic.
\end{enumerate}

The first two assumptions are the main reasons why the sensitivity criterion fails in 
the case of  renormalization. For the first assumption, because for a 
fine-tuning problem associated with  renormalization,  
it is usually hard to define an upper bound of an input parameter
naturally. The sensitivity criterion implies the value \(x_0\) is the upper bound of 
the input parameter \(x\), and ignore the possibility of \(x \ge x_0\). Obviously, 
P. Ciafaloni did aware this problem, and change the definition of the sensitivity from 
its original version (Eq.~(\ref{eqx4})) to the modified one(Eq.~(\ref{eqx9})), which means it is assumed 
that the Lagrangian parameter \(x\) uniformly distributed from \(0\) to the experimentally
allowed maximum value \(\Delta x\). Certainly this is a step ahead toward the correct prescription
but we also have difficulties to choose a specific experimentally allowed maximum value.

The second assumption means that what the sensitivity \(c\) tells us is always the average 
fine-tuning level
of an observable parameter \(y\) to be found in the region \(0 \le y \le y_0\).
It does not reflect the exact fine-tuning level at \(y=y_0\).
This problem will be even worse in renormalization. Because
for the fine-tuning problems related with  renormalization, 
parameter ranges \([0, y_0]\) and \([0, x_0]\) are related mathematically.
According to  our previous 
discussion, the sensitivity parameter can only reflect the contribution from the 
anomalous dimension. It will give a wrong answer, although the big ratio of 
\(y_0/x_0\) is the origin of the fine-tuning problem.
These assumptions especially
the second one will have severe consequences under some special conditions.

For the third assumption, because Eq.~(\ref{eqx10}) is only correct when \(y=f(x)\) is monotonic,
thus the sensitivity criterion also implied  the relation between 
the input and the output is monotonic, which means we can't find an example that both input 
parameters \(x_1\) and \(x_2\) will eventually lead to  a same output parameter \(y\), certainly,
it is not true. Because although the relationships between parameters linked by most renormalization group equations
are monotonic, while most mixing cases are not. One output parameter  usually
corresponds to two input parameters. For example, mixing of \(M_Z\) mass and \(M_W\) mass, 
mixing of CP-even Higgs masses in Supersymmetric Standard Models, and mixing of fermionic masses etc. 

Take the mixing of \(M_Z\) mass in MSSM model as an example, calculate the \(M_Z\) mass 
at the initial condition (at grand unification scale) \(m=200\mbox{~GeV}, M=40\mbox{~GeV}, \tan\beta=18 \), 
and gradually reduced mass \(m\), then we will find the non-monotonic relationship between 
grand unification scale variable \(m\) and weak scale variable \(M_Z\).

Obviously, \(M_Z\) is not a monotonic function of \(m\), for example, if  weak scale mass
\(M_Z\) is around \(80GeV\), there are two grand unification scale parameter regions that
can contribute this result, one is around \(m\) is \(80\) GeV, the other region is around
\(m\) is \(150\) GeV. Barbieri and Giudice's definition only counted one region's contribution
, it will overestimate the naturalness level.
We should count all possible GUT scale parameters contributions.

\section{Dimensional effect}

It is meaningless to compare two different parameters with completely different units. 
In order to do so we need first convert them to a comparable format. 
The sensitivity criterion uses the relative variation to convert a dimensionful parameter \(y\)
to a dimensionless format  \(\delta y/y\).  It looks like the problem
of comparing parameters with different units has been solved, but
the examples given out by G. Anderson et al\cite{GWA}(Eq.~(\ref{eqx5})), 
and P. Ciafaloni et al\cite{CS}(Eq.~(\ref{eqx7}))
remind us it still has problems when comparing parameters
with different units.

Generally, if we only consider the contribution from the naive dimensions and 
ignore the small corrections like the anomalous dimensions, physical parameters
can be categorized into two types: scale invariance conserved 
and the scale invariance broken\cite{wilson2}. The scalar mass of $\phi^4$ model is an example
of scale invariance conserved. As the energy scale \(\Lambda\) increases, the value of the
parameter will also be increased as an exponential function \(\Lambda^{\alpha}\)
(where \(\alpha\) is the naive dimension);
While the fermion masses is an example of the scale invariance broken, where the value of the 
parameter will not increase as an exponential function of the energy scale \(\Lambda\).
We can find that if the naive dimension of a parameter is not equal to zero, 
those parameters with  scale invariance conserved
will have the fine-tuning problem in renormalization, because the values of these parameters
will blow up quickly as the energy scale increases.

If the scale invariance is conserved,  for two  parameters 
\(\tau\) and \(h\) with different naive dimensions, In renormalization, if we change 
the energy scale  \(\Lambda\), then these two parameters \(\tau\) and \(h\)
will have the following relations: 
\begin{equation}
\tau\approx\Lambda^{\alpha}\tilde{\tau}
\label{eqx14}
\end{equation}
\begin{equation}
h\approx\Lambda^{\beta}\tilde{h}
\label{eqx15}
\end{equation}
where \(\alpha\) and \(\beta\) are the corresponding naive dimensions.
Because of the renormalization, 
the energy scale \(\Lambda\) will connect these two parameters together,
even they may not have any other mathematical relation.
If we calculate the sensitivity of \(h\) to the variation of 
\(\tau\),  because:

\begin{equation}
\frac{\partial h}{\partial \tau}\approx -\frac{\beta}{\alpha}\frac{h}{\tau}
\label{eqx16}
\end{equation}
we can immediately calculate the sensitivity \(c\) is \(\beta/\alpha\).
This sensitivity is a consequence of comparing two parameters with 
the different naive dimension.
If both \(\tau\) and \(h\) has the same naive dimension, then the sensitivity 
will be one, thus the dimensional effect will disappear. But when they have different canonical
dimensions, this factor will not be one, it will greatly affect the sensitivity.

This effect is known as scaling effect in statistical physics, which exists anywhere 
when two parameters have different dimensions. Obviously,
it has nothing to do with the fine-tuning. In the definition of the sensitivity
parameter \(c\), in order to compare parameters with different dimensions, 
the relative variation \(\delta y/ y\) was introduced to remove the effects of 
different dimensions,
but based on Eq.~(\ref{eqx16}), 
we find the dimensional effect is a nonlinear effect, can not be eliminated by 
the relative variation \(\delta y/ y\). When considering the fine-tuning problems 
in renormalization, the effect of different naive dimension \(-\beta/\alpha\)
is still there. This is the reason why the sensitivity criterion failed in the
examples given by G.W.Anderson et. al. and P.Ciafaloni et. al. 

If one parameter \(\tau\) has a marginal naive dimension, or when the scale invariance 
is not conserved, then we can not use the 
above argument, we must consider the 
higher order term. take the coupling constant renormalization as an example:
\begin{equation}
\tau\approx(1+\alpha\tilde{\tau}\ln\frac{1}{\Lambda})^{\delta}\tilde{\tau}
\label{eqx17}
\end{equation}
where \(\delta\) will be \(1\) for parameters with zero naive dimension (for example,
the gauge coupling constants), or a non-zero number (for example, the fermion masses).
Corresponds to Eq.~(\ref{eqx16}), we have:
\begin{equation}
\frac{\partial  h}{\partial \tau}\approx\frac{\beta}{\alpha\tau\delta}\frac{h}{\tau}
\label{eqx18}
\end{equation}

The corresponding sensitivity \(c\) will be \(\beta/\alpha\tau\delta\). Obviously, this
factor can be very large. For example, the problem of the Z boson mass  
discussed previously, even it is a fermion with zero dimension, we still find 
large sensitivity when comparing with the coupling constant, this is due to the 
factor \(\delta\) here is not zero. Once again, this reminds us that the dimensional
effect is a nonlinear effect, the relative variation \(\delta y/y\) used in the 
sensitivity criterion can only remove linear effect. Thus even the unit has been removed by using \(\delta y/y\),
its effect is still there. The factor \(\beta/\alpha\tau\delta\) is the contribution of the dimensional effect 
to the sensitivity.

For convenience we define a dimensional effect factor:
\(\Delta=\beta h/\alpha\tau\) or
 \(\beta h/\alpha\delta\tau^2\) for later reference.

\section{Definition of the new criterion}

As we discussed in the previous sections, the traditional sensitivity criterion
fails mainly for two reasons, first,  it uses the total probability instead of the
probability density, which can only reflects the average sensitivity of the whole
parameter range. For the fine-tuning problems in renormalization, 
because the total probability is not greatly changed even when the parameter is fine-tuned,
using the total probability will give the wrong result.
Second, because the sensitivity criterion does not include the contribution from the dimensional
effect, so it fails when comparing parameters with different dimensions.

Generally, there are two types of the fine-tuning problems, either because of the 
parameter space greatly shrunk from the input to the output (in renormalization), 
or because of the parameter range for both input and output are extremely large(in mass mixing).  

For the first type, as we have discussed,  it is not a good idea to 
use the total probability to calculate the fine-tuning level, 
because a large naive dimension is the main reason 
why we have the fine-tuning problems in renormalization, while the constant part of the dimension 
will not change the total probability, and this part happens to be the main reason of the
fine-tuning. So we need a criterion that can count the contribution 
from the whole naive dimension. Besides, for this type we also need to consider the dimensional effect when 
comparing parameters with different mass dimensions.

While for the second type, because it happens at the same energy scale. For the sensitivity criterion,
we only need to face the problem of using the total probability. 
Using the total probability in some sense does reflect the fine-tuning level qualitatively,
although strictly speaking it only gives an average level of fine-tuning in a certain range. 
So it is more important to find a new fine-tuning criterion for the first type.

Before proceed to propose a new prescription, we have to solve two problems.
First, because we don't know the distribution function of the Lagrangian parameters at the grand 
unification scale. Without any experimental evidence, we have to assume that these 
Lagrangian parameters are all evenly distributed.

Second, how to choose the maximum allowed values of these parameters is another important problem.
because this will also greatly affect the final result. In the sensitivity criterion,
it is implied that the parameter value we want to measure the fine-tuning level is
the maximum allowed value. Later P. Ciafaloni et al. modified it to the experimentally
allowed range\cite{CS,BR,BS,GRS,RS}. Certainly, for the fine-tuning problems existed in mixings
(second types), because the parameter range is not specified by the mixing mechanism 
itself, so we can choose a maximum allowed value (either experimentally or theoretically)
as the maximum parameter range.
While for the first type, unlike the second type, 
the maximum allowed parameter range is usually given by the renormalization itself. Once
the weak scale is given, the maximum parameter value at the grand unification scale 
will be controlled by the naive dimension. For example, for \(\phi^4\) model, we need to 
assume at the grand unification scale, the maximum allowed mass is around \(10^{18}\) GeV,
while for a fermion mass, we only need to assume the maximum allowed mass is around couple 
hundred GeV. Now if we want to compare a \(\phi^4\) model scalar mass with a fermion mass, 
which maximum allowed mass we should choose?  \(10^{18}\) GeV?  or couple hundred GeV?
Either way will be a disaster for any fine-tuning criterion. 

In mathematics, sensitivity to initial conditions can be measured quantitatively in different ways,
for example, by Lyapunov exponent~\cite{predrag}, which measures the separation of two neighboring 
trajectories in the phase space.  Using the language of the probability theory, it corresponds to measure
the variation of the probability distribution. 
This problem is almost identical to the fine-tuning problems existed in renormalization,  
we can borrow its idea to define a new fine-tuning criterion in renormalization.

In order to solve the problems we discussed previously, 
we can not compare the relative variation \(\delta x_0/\delta x\) with \(\delta y_0/\delta y\), 
instead, we need to use \(\delta y/\delta x\) directly. 
Here the factor \(\delta y/\delta x\) means the ratio between the variation \(\delta x\) and
the corresponding variation \(\delta y\). Just like the problem of sensitive to initial conditions
in mathematics, the Lyapunov exponent (corresponds to the dimension in physics) decides the 
final result, we don't need to give a maximum allowed value. For the fine-tuning problems in renormalization,
use \(\delta y/\delta x\) directly, will solve all the problems in the sensitivity criterion.

Certainly, we need to consider the dimensional effect, 
if two parameters \(x\) and \(y\) doesn't have any other mathematical relation except 
linked by the energy scale \(\Lambda\) in renormalization,
because of the dimensional effect, \(\partial y/\partial x\)
will be equal to the dimensional factor \(\Delta\) we defined earlier. If both parameters 
have the same naive dimension, then the factor \(\Delta\) becomes one. This factor is a kind of
background probability density which should always be subtracted when consider the fine-tuning problems. 
After subtracted the factor \(\Delta\), the remaining part of \(\delta y/\delta x\) is due to the fine-tuning 
mechanism.

Because both  \(\delta y_0/\delta y\) and the dimensional factor \(\Delta\) has the same 
mass dimension, in order to have a dimensionless fine-tuning measurement,   refer to 
the definition of the Lyapunov exponent,  if we take the dimensional effect factor \(\Delta_0\) as 
the initial separation of two neighboring renormalization group orbits, then the separation 
between these two orbit can write as the function of \(t\):

\begin{equation}
\frac{\delta y}{\delta x}=\Delta_0 e^{\lambda}
\label{eqx19}
\end{equation}
Where \(\Delta_0\) is the dimensional factor at the grand unification scale.
Thus the factor \(\lambda\) here is always dimensionless.

According to Eq.~(\ref{eqx11}), \(\partial y/\partial x\) is identical to the probability density,
the dimensionless factor \(\lambda_0\) at the grand unification scale reflects the 
background probability density. So the factor \(\lambda\) represents
how difficult to choose a specific value at the weak scale. 
Using the language of the probability theory, Eq.~(\ref{eqx19}) means after 
considered the dimensional effect, if at the weak scale one parameter has 
a fluctuation in a small area, then  at the grand unification scale, 
the probability of the corresponding parameter still within this small area 
is \(\exp{\lambda}\). Because we didn't use the relative variation in the beginning, so
clearly, this prescription won't have any problems the sensitivity criterion has. 

Just like the Lyapunov exponent, the factor \(\lambda\) here corresponds to the naive dimension. 
If a parameter has a big naive dimension, for example, the scalar mass  of \(\phi^4\) model, we will have a 
large positive \(\lambda\), which means as the energy scale increase, the scalar mass increases rapidly
and will have a fine-tuning problem. While for a fermion masse, usually, we will have a small
negative \(\lambda\), which means as the energy scale increases, the fermion masses decrease
slowly thus won't have the fine-tuning problem.

Now if considering the non-monotonic propriety, according to the probability theory, we
should divide the whole parameter region into several monotonic sections, and sum up the corresponding
probability densities. Based on this, we can modify the previous expression to the following form:
\begin{equation}
\lambda=\ln\sum\bigg|\frac{1}{\Delta_0}\frac{\delta y}{\delta x}\bigg|
\label{eqx20}
\end{equation}
Which means if it is not monotonic, we divide and sum up the exponent over all monotonic regions.

\section{Definition the maximum tolerated $\lambda$ }

Although the problem of the fine-tuning criteria in high energy physics is somewhat
similar to the problem of using Lyapunov exponent to judge whether a 
nonlinear system is sensitive to initial conditions or not, these two situations also have  
important differences. 

In nonlinear physics, if the Lyapunov exponent is negative, then the phase
space of a system shrinks as time increases, it is obviously not initial condition sensitive.
If the Lyapunov exponent is positive, then the system is
initial condition sensitive and thought it is chaos.
Similarly, when we consider the fine-tuning problem in high energy physics, 
if \(\lambda<0\), for the same reason, we can easily classify the system as not
initial condition sensitive or not fine-tuned. But for the cases with
\(\lambda>0\), the situation is a little more complex. This is because for systems 
in nonlinear physics, 
the time variable \(t\) can go to infinity while in high energy physics,
the running parameter \(t\) can not go beyond the grand unification scale, 
which is around \(38\). So for the situations with
small positive \(\lambda\), even the corresponding variation in the weak scale is
a little less than the variation in the grand unification scale, it still can be 
thought as not initial condition sensitive, or not fine-tuned. So 
a small positive \(\lambda\) should not be considered as fine-tuned.

In probability theory people usually define the probability \(P\le 0.05\) as the limit of
small probability events, although this is a more strict 
condition than the
sensitivity less than \(10\) criterion (which means the \(P\le 0.1\)), 
because  the criterion of \(P\le 0.05\) is widely used in probability theory,
so here it is used to define an upper limit for the fine-tuning level. 
Suppose fine-tuning occurs when \(P\le 0.05\),  or \(\exp(\lambda )\le 1/0.05\).
we immediately have the upper limit of the \(\lambda\) is 3.
According to this definition, all parameters with \(\lambda \le 3\)  will be safe
and not fine-tuned, and if \(\lambda > 3\) which means it is less than 5\% of chance
to have this weak scale value thus it is quite impossible. Not like Barbieri and Giudice's
naturalness \(c\approx 10\) cut-off, which doesn't have any physical meaning, our method
gives a clear physical meaning of the naturalness cut-off.

\section{Examples}

\begin{enumerate}

\item The \(\Lambda_{QCD}\) problem:
The sensitivity criterion gives \(c\gtrsim 100\), it is fine-tuned, which is wrong.
For our new criterion, it is not difficult to calculate that
\(\lambda=(\ln\frac{\Lambda_{QCD}}{gM_P})\), which is far less than 3. It is not fine-tuned,

\item For non-monotonic problem:

for non-monotonic case \(M_z\), if \(M_z=89.05\) GeV, which corresponds to \(m=188.5\) GeV and
\(m=65.1\) GeV at the grand unification scale,  we calculated that sensitivity 
\(c=0.499\) and \(c=0.910\) respectively. while \(\lambda\) for \(M_z=89.05\) GeV is \(-1.748\).
Although both of them don't have much difference, while if we use the fine-tuning level to
set the allowed parameter range to guide our new particle search, then the contour will be wrong.

\item \(\Phi^4\) model scalar mass problem:

for \(\Phi^4\) model scalar mass,
which has engineering dimension equals to \(1\), sensitivity \(c=1<10\), which is wrong.
While our new criterion
\(\lambda=1+\frac{g^2_{GUT}}{32\pi^2}\), greater than \(3\).

\end{enumerate}

All these examples show that our new naturalness criterion can solve all problems
the sensitivity criterion has.

\section{Conclusion}
The sensitivity criterion is widely used in measuring the level of fine-tuning,
although it is not reliable.
In this paper we  investigated the mathematics  behind the fine-tuning problems,
categories these fine-tuning problems into two types: the first and the second type.
Then based on the probability theory, we analyzed the problems existed in the 
sensitivity criterion, and find three assumptions that widely ignored.
By analyzing the high sensitivity problems of \(\Lambda_{QCD}\) and Z-boson mass, 
we  found  the reason why the sensitivity criterion failed is because of 
the dimensional effect, because of the dimensional effect the sensitivity is much larger than the level of 
real fine-tuning. This effect should be removed when considering fine-tuning problems in
renormalization. Based on these analysis, we found the traditional sensitivity criterion 
can not be used to measure the fine-tuning level for the first type, although it 
did reflect the fine-tuning level qualitatively for the second type. 
Based on the probability theory, we proposed  a new criterion that can solve all these problems,
and verify our new criterion with various fine-tuning problems.

\acknowledgements{ The author would like to thank Professor Gregory W. Anderson for 
the guidance and inspiring discussions.
This work was supported in part by the US Department of Energy,
Division of High Energy Physics under grant No. DE-FG02-91-ER4086.}

\end{document}